# X-RIS: A Study of the Principles and Applications of X-Shaped RIS

Xiaocun Zong, Binchao Zhang, *Member, IEEE,* Fan Yang, *Fellow, IEEE,* Shenheng Xu, *Member, IEEE,* and Maokun Li, *Member, IEEE*

*Abstract*—**This paper analyzes the working principle of X-Shaped reconfigurable intelligent surface (RIS) in detail and reveals the different types of RIS that can be designed based on this structure. Combined with the design examples using this structure in the currently published articles, this paper summarizes and organizes them, and finally, based on this X-Shaped structure, this paper explores some other possible designs, which reflects the potential of the design versatility of the X-RIS structure.**

## I. INTRODUCTION

RECONFIGURABLE intelligent surface (RIS) is a concept of programmable array consisting of a large number of periodically arranged elements, and is a new kind of phased array. According to the different feed positions, it can be divided into reflect reconfigurable array (RRA) and reconfigurable transmit array (RTA). Thanks to the spatial feeding and integrated phase-tuning technique, RIS has a simple and low-cost structure compared to conventional phased arrays [1]-[3], and thus have become more and more popular in academia and in various civilian and military applications.

The RIS element is adjustable, and the beam scanning function in different directions can be realized by controlling FPGA. In the RIS design, the key to beamforming is to assign the required phase value to each element. The most popular method to realize phase distribution is using PIN diodes [4], [5], the PIN is loaded into the element patch or the phase-shifting structure, and different working modes can be formed

This paragraph of the first footnote will contain the date on which you submitted your paper for review, which is populated by IEEE. It is IEEE style to display support information, including sponsor and financial support acknowledgment, here and not in an acknowledgment section at the end of the article. For example, "This work was supported in part by the U.S. Department of Commerce under Grant 123456." The name of the corresponding author appears after the financial information, e.g. *(Corresponding author: Second B. Author).* Here you may also indicate if authors contributed equally or if there are co-first authors.

The next few paragraphs should contain the authors' current affiliations, including current address and e-mail. For example, First A. Author is with the National Institute of Standards and Technology, Boulder, CO 80305 USA (e-mail: author@ boulder.nist.gov).

Second B. Author Jr. was with Rice University, Houston, TX 77005 USA. He is now with the Department of Physics, Colorado State University, Fort Collins, CO 80523 USA (e-mail: author@lamar.colostate.edu).

Third C. Author is with the Electrical Engineering Department, University of Colorado, Boulder, CO 80309 USA, on leave from the National Research Institute for Metals, Tsukuba 305-0047, Japan (e-mail: author@nrim.go.jp).

by controlling the switch of the PIN. According to the number of phase states, it can be divided into 1-bit, 2-bit and even 3 bit, then we call this phase quatization.

In the RIS design, two very important components are the radiation patch and the phase shift structure. The patch is used to receive electromagnetic waves and radiate the phase-shifted electromagnetic signals. There are various radiation patches in the existing literature, such as "U-shaped"[6].[7], "square-shaped"[8],[9], and"circular-shaped"[10]. In theory, any shape can radiate, but in order to better match and radiate bandwidth, some more detailed designs are usually carried out. There are also various solutions for the phase shift structure, such as different resonant states produce different reflection phases[11]-[13], control the length of the microstrip line to use the path difference to produce different phases[14],[15], etc. How to better combine the radiation part and the phase shift part is also a problem that needs to be considered in the design process.

The X-Shaped patch perfectly solves the problems mentioned above, and it has been used in a large number of RRA/RTA designs due to its simple and symmetrical structure, mature optimization methods, the diverse radiation states and the advantages of integrated "radiation-phase shift" control. This structure includes a metal ground, dielectric substrate layer, a pair of cross-placed metal strips, and extension structures at the end of the metal strips. It also includes four PIN switches. It will have slight deformations depending on different application scenarios. What is even more surprising is that this structure shows different effects when placed parallel to the electric field and when placed at a 45° angle. This feature can also bring different inspirations to our design. When the electric field is incident along the diagonal direction of the element, different resonant states can be formed and various designs can be performed [16]-[25]; When the electric field is incident along the edge of the element [26]-[46], a polarization conversion effect can be formed, and a broadband effect can be formed through various optimizations; combining different states will produce other functions; even combining patches in two directions can produce more unexpected effects. This structure can not only perform phase shift processing on electromagnetic signals, but also can form high-order bit phase quantization numbers based on the original simple structure by controlling the switch state and element angle. If you add some other structures to this structure, you can get even more design inspiration. At the same time, the patch can also radiate, truly achieving an integrated functions of high-order quantization and radiation. Therefore, it is necessary to conduct in-depth exploration and detailed summary of this X-Shaped patch structure.

This paper will conduct a detailed theoretical analysis of the working principle of this X-Shaped RIS structure, then review and organize different designs. Finally, several new design methods are proposed based on the theoretical analysis of this paper.

## II. BASIC STRUCTURE AND PRINCIPLES

The X-Shaped RIS has a symmetrical structure and a simple composition. It usually includes a metal ground, a pair of cross-placed metal strips patch, PIN switches on the metal strips, and extension structures at the end of the metal strips, which is generally used to expand bandwidth. If this structure is used in a RTA, the metal ground is replaced by another radiate patch. The relative position relationship between the electric field and the element varies in different application scenarios, as shown in Fig.1.

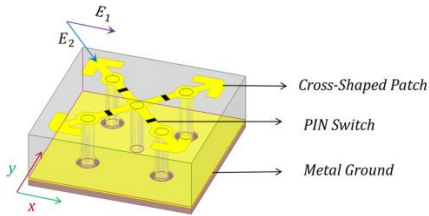

Fig. 1.   The general structure of X-Shaped patch RRA.

### A. The principle of the electric field is incident along the edge of the element

When the electric field is incident along the edge of the element, just like $E_1$ in Fig1, the patch plays a role in polarization conversion. If the electric field is incident along the $x$ direction, the electric field can be decomposed into two components in the $u$ and $v$ directions. The current components in the two directions produce a 180° phase difference after flowing through the patch, and after synthesis, they become the electric field in the $y$ direction, the principleis shown in Fig.2, this is the specific principle of polarization conversion.

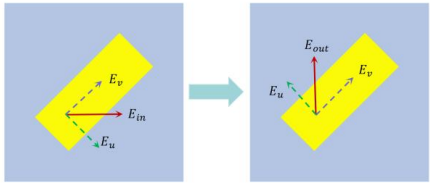

Fig. 2. The principle of polarization conversion by placing patch diagonally

According to the mirror law, when the two patches are placed along ±45°, the electric fields emitted by the two patches will produce an electric field phase difference of 180°. Based on this principle, we can construct a 1-bit RRA element, combining the two patches and controlling the orientation of the patches by loading 4 PIN diodes, as shown in Fig.4: when PIN diodes #1 and #2 are in the OFF state and #3 and #4 are in the ON state, it is state 1; when PIN diodes#1 and #2 are in the ON state and #3 and #4 are in the OFF state, it is state 2.

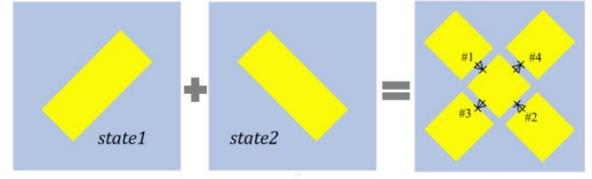

Fig. 3. Schematic diagram of the principle of a 1-bit RRA element

By optimizing the size structure of the element, multi-mode resonance can be used to achieve broadband design. And it is not difficult to find that as long as it is a rectangular strip structure placed along the diagonal and forming an angle of 45° with the electric field, it can form a polarization conversion effect. By using the mirror principle, placing two patches at ±45° will form a phase difference of 180°. It is just that the polarization conversion working frequency produced by different lengths is different. The element under this electric field incidence mode can form three polarization conversion effects at different frequency according to the number of PIN switches opened, that is, the B4, B5, and B6 states in Fig.4.

In addition to the polarization conversion effect, there are still three resonance modes under this incident electric field mode according to the different equivalent resonance lengths. When all PIN switches are in the off state, the resonance state B1 is formed; when the two adjacent branch PIN switches are turned on and the equivalent resonance length is half the side length, it is the resonance state B2; when all PIN switches are turned on, or the two adjacent branch switches are turned on to form an equivalent resonance length of the same length as the side length, this is the resonance state B3, which enriches the design diversity. As shown in Fig.4, all the design possibilities are plotted and the equivalent resonant size can be seen.

### B. The principle of the electric field is incident along the diagonal of the element

When the electric field is incident along the diagonal of the element, just like $E_2$ in Fig.1, the effect at this time is almost the opposite of the diagonally placed mentioned above. The element mainly works in the resonant mode. Different resonant states can be formed according to the number of PIN diodes turned on. At the same time, this element also has a polarization conversion state, which provides more possibilities for design diversity.

According to the relative position relationship between the cross metal strips and the incident direction of the electric field, different resonance principles can be formed. If the metal strip branch in the switch-on state is perpendicular to the incident electric field, then no matter how complex the structure of the branch perpendicular to the electric field is, they are equivalent to the same resonant state, because the vertical branch does not resonate, and this is state A1. When only one PIN switch is turned on in the branch along the electric field direction, no matter how many PIN switches are turned on in the vertical branch direction, the resonant state at this time is equivalent, which is the resonance principle of A2. Similarly, when two PIN switches are all turned on in the branch along the electric field direction, no matter how many PIN switches are turned on in the vertical branch direction, the resonance state at this time is equivalent, which is the resonance principle of A3.

Fig.4. All possible states of the electric field is incident along the edge of the element.

In addition to the three resonant states mentioned above, when the electric field is incident along the diagonal direction, a polarization conversion effect can also be formed. When two adjacent branch switches are turned on, the current will change the direction of flow, forming a polarization effect. This is the principle of A4. Then we draw all the possible states and corresponding principles in Fig.5.

## III. REVIEW OF DESIGN CASES

From the analysis in section II, we can see that by changing the incident direction of the electric field, the 32 forms of the X-Shaped element structure can form 10 resonance or polarization conversion modes. Combining these modes, a variety of RA and TA designs can be designed, and elements with different bit numbers can be flexibly designed.

When the electric field is incident along the edge of the element, if only two opposite switches are open (B5), a large number of works have been designed using this principle, that is, the commonly used 1-bit polarization conversion RIS element, references [26],[27],[29]-[34] made a polarization conversion broadband RIS, this principle can be used to design both RA and RA, both of which can achieve ultra-wideband, which is also the most widely used application of this structure, even some design can achieve bandwidth expansion of three times its own center frequency; what's more, the RIS designed using this polarization conversion principle can also be used to generate orbital angular momentum (OAM) beams, references [28] and [35] use this principle to design 1-bit TRA and RRA to generate OAM; there are also a lot of articles using this structure to achieve RCS reduction, [37]-[43] design RIS by exploiting polarization conversion states to suppress the scattering of metal objects by the combination of scattering and reflection, which has significant potential in the applications of antenna designs or stealth technology fields; besides, this design of RIS has also made achievements in the field of communications, reference [44] proposes a broadband metasurface-based wireless communication system that can actively adapt to multiple users located at versatile directions through joint modulation of digital signals in the time domain and wave scatterings in the space domain; moreover, this structure can also be used to design 1-bit transmission array by slightly modifying it, references [19] and [20] designed TA based on this structure to achieve polarization conversion, reference [19] realized a wideband 1-bit filtenna-to-filtenna cross-polarization converter by using multimode resonance, reference [20] used this structure to construct a non-reconfigurable transmission array that can achieve circular-to-linear polarization conversion. Besides, reference [36] adjusted the patch size, and made a pair of dual-frequency reconfigurable polarization conversion reflectarray, which is capable of generating two beams at X- and Ku- bands.

Fig.5. All possible states of placing the element parallel to the E field.

In addition to the commonly used B5 mode, other modes can also produce different effects. In the references [16], a 1-bit independently controllable dual circular polarization RRA was designed by using two resonant states and two polarization conversion states (A2 and A4); reference [18] achieved 1-bit independent control of dual-linear polarization by introducing additional bias, then realizing four states (A1 and A2); in the references [21], this article designs a dual-polarized 2-bit RRA, by introducing an additional PIN diode (A1, A2 and A3); reference [45] uses two polarization conversion states and two resonance states (B1, B3 and B6) to design a 2-bit RRA element by combining these four states; reference [46] uses four states: all PIN diodes are off, two states where only one PIN is on, and one state where two PIN diodes are on (B1, B4 and B5), realized the 1-bit phase resolution in the two polarization direction.

There are also papers that use a combination of two incident directions for design. In reference [24], the authors designed a non-reconfigurable 2-bit RRA, using four states (A1, A3 and B5), which is shown in the Fig.6, and a vector beam modulator is achieved by combining orthogonal polarizations and orbital angular momentum modes. In reference [25], a 2-bit coding metasurface for ultra-wideband and polarization insensitive RCS reduction is designed using the same four states.

Considering the number of phase quantization bits, the functional classification implemented, and the corresponding design principles, Fig.6 provides a detailed classification and review.

## IV. NOVEL DESIGN EXAMPLES

Based on the analysis of the above principles and the existing works, we explored other possible designs based on this X-Shaped structure. Out of curiosity, what will be the effect of opening another branch on the basis of the polarization conversion effect of B5? That is, state B6. It can be predicted that there must be a certain frequency band that can achieve 100% polarization conversion. At the same time, there will also be a certain frequency band where polarization cancellation occurs due to the opposite directions of the two cross currents, thereby having a certain impact on the reflect amplitude and phase. Polarization cancellation brings loss, but the reflection phase can also be adjusted based on this. We use the two states B5 and B6 to realize a variable bit number reflectarray of 1-bit UWB and 2-bit narrowband, we call this "bit reconfigurable".

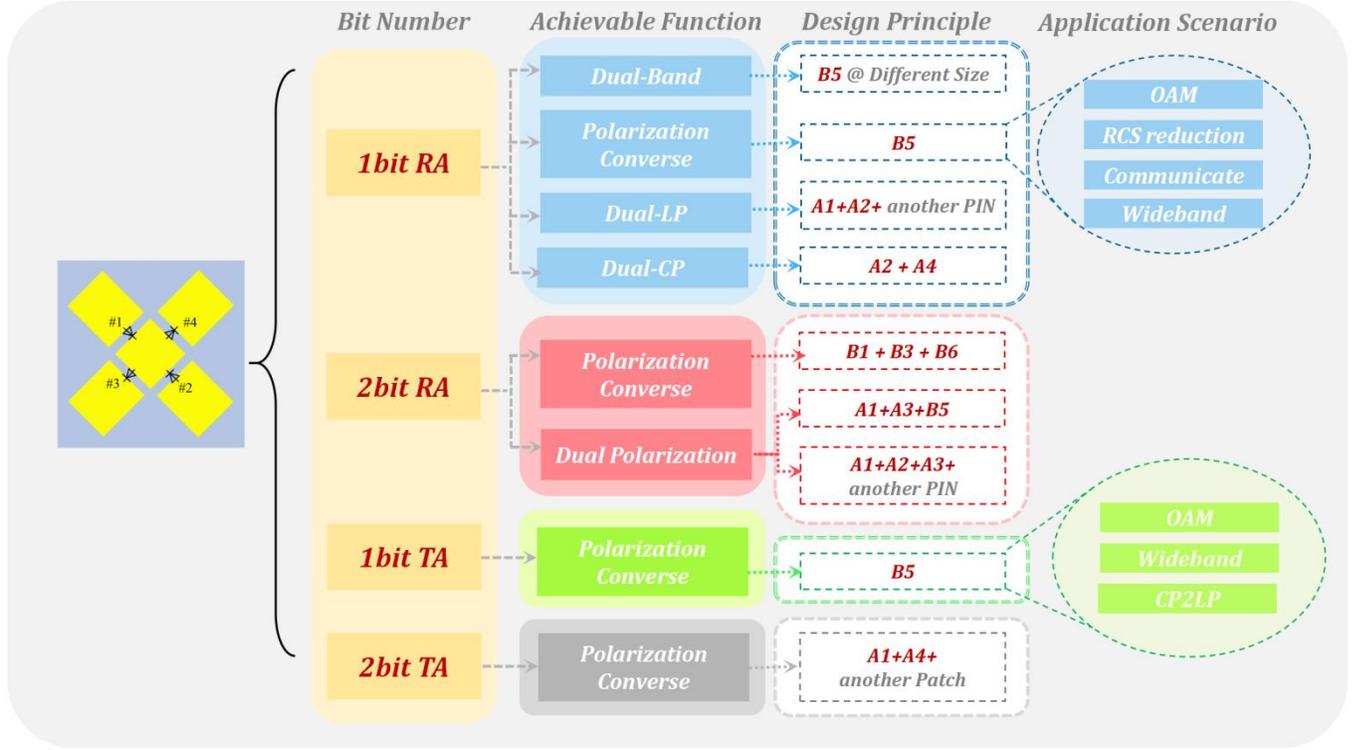

Fig.6. Based on X-Shaped structure some existing works

Besides, inspired by reference [36], which uses two different sizes of elements to achieve dual-band function ,but we want to use one element to achieve dual-band function. Based on this motivation, by using the change of patch length design, different polarization conversion states can be combined to design a dual-band ultra-wideband 1-bit element. At this point, four switches need to be controlled independently to achieve dual-band conversion.

The above two examples are both polarization conversion antennas. According to the principle analyzed above, by introducing an additional PIN switch, we can realize a co-polarization reflectarray with four resonant states.(A1, A2, A3 and another state). We have reviewed the possible designs, as shown in Fig.7.

In short, this X-Shaped RIS structure can achieve a variety of designs and functions, and there are still many possible designs worth exploring.

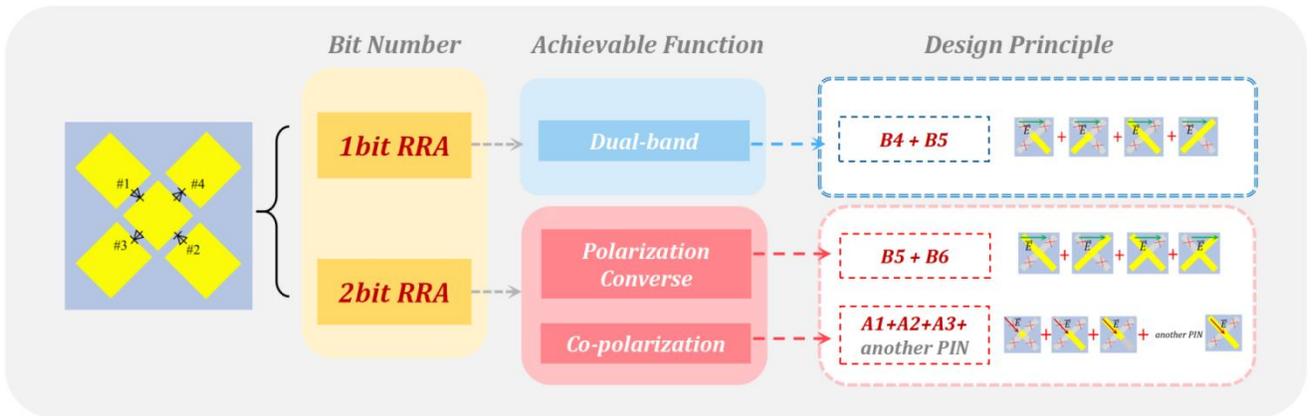

Fig.7. Based on X-Shaped structure our proposed possible design.

## IV. CONCLUSIONS

This paper systematically reviews the X-Shaped RIS, including the principle analysis and overview of the existing works, they are classified according to the principle, and finally, based on the analysis of the X-RIS principle, we propose several possible designs, which further broaden the diversity of this structure.